\documentclass[doublecol]{epl2}

\usepackage{graphicx}
\usepackage{amssymb,amsfonts,amsmath}

\title{Storage capacity of phase-coded patterns in sparse neural networks}

\author{S. Scarpetta\inst{1,2} \and F. Giacco\inst{1,2} \and A. de Candia\inst{3,4,2}}

\institute{
  \inst{1} Dipartimento di Fisica ``E. R. Caianiello'', Universit\`a di Salerno, Italy\\
  \inst{2} INFN, Sezione di Napoli e Gruppo Collegato di Salerno\\
  \inst{3} Dipartimento di Scienze Fisiche, Universit\`a di Napoli Federico II\\
  \inst{4} CNR-SPIN, Unit\`a di Napoli\\
}

\abstract{%
We study the storage of multiple phase-coded patterns as stable dynamical
attractors in recurrent neural networks with sparse connectivity.
To determine the synaptic strength of existent connections and store the phase-coded patterns,
we introduce a learning rule inspired to the spike-timing dependent plasticity (STDP).
We find that, after learning, the spontaneous dynamics of the network replay one of the stored dynamical patterns,
depending on the network initialization.
We study the network capacity as a function of topology, and find that a small-world-like
topology may be optimal, as a compromise between the high wiring cost of long range
connections and the capacity increase.
}


\begin{document}

\maketitle

The capacity to memorize and recall multiple items
of information is fundamental to normal cognition.
Recent researches suggest that brain operates as a complex nonlinear dynamic system,
and synchronous and phase-locked oscillations may play a crucial role in information
processing, perception, memory, and sensory computation
 \cite{science, Singer99, Fries2007, buz, MillerPNAS}.
There is increasing experimental evidence
that information encoding  may depend on the temporal
dynamics between neurons, namely, the specific phase alignment
of spikes relative to rhythmic activity across the neuronal
population (as reflected in the local field potential)
 \cite{8,burges,zugaro-cn,odor, Montemurro,12,MillerPNAS,panzieri}.
Indeed phase-coding, that exploits
 the precise temporal relations between the spikes
 of neurons, may be an effective general
 strategy to encode information in the cortex
\cite{NC,mate,Timme2006,PREYoshioka,y,Bush,masquelier,nature}.

The importance of precise timing of the
neuron activity is also suggested by the rule that
controls the potentiation or depression of synaptic strengths,
namely the spike-timing dependent
plasticity (STDP) \cite{markram,biandpoo}, based on the precise order
and time interval between
the pre- and post-synaptic spikes, on a window of tens of milliseconds.
This kind of plasticity, with acute sensitivity to temporal order,
 has been demonstrated  in various
neural circuits over a wide spectrum of species,
from insects to xenopus laevis frog, rodents and
humans \cite{caporale,sjostrom}.

The STDP is strongly asymmetric in the order of arrival of pre-
 and post-synaptic spikes,
{usually} determining a potentiation
in the case of causal order (pre-before-post), and depression in the
reverse order. This {temporal asymmetry} results
in asymmetric connections between neurons,
that is a crucial ingredient
to give rise to dynamical patterns, as opposed to static patterns
characteristic of symmetric
connections as in the Hopfield model.

Another crucial feature of the network is its connectivity and topology.
Whereas all-to-all, or random, wiring is usually assumed in many models,
this is possible in a tissue culture involving dozens of neurons, but it becomes
less and less feasible when millions of neurons
are involved, owing to space and energy supply limitations.
The topology of the brain connectivity has been designed by
evolutionary goals as a compromise between the ability to achieve
complex dynamical functions and  physical constraints and costs minimizations
\cite{17,15}.

In the last decade, there has been a growing interest in the study of the topological
structure of the brain network.
There is increasing evidence that the connections of neurons in many areas of the nervous
 system have complex topology,
such as a small world topology \cite{watts-strogats}, highly
connected hubs and modularity \cite{bullmore}.
Up to now, the only nervous system to have been comprehensively
mapped at a cellular level is the one of Caenorhabditis elegans,
and it has been found that is has indeed a small world
structure \cite{watts-strogats,latora}.
The same property was found for functional
and anatomical connectivity,
in different animals and areas of the brain \cite{singer,sporns1,bullmore}.

In this paper, we focus on the functioning of the network as a
dynamical associative memory,
that is on the ability of the network to memorize and recall multiple dynamical patterns,
where each pattern is characterized not by a set of binary states of the neurons,
as in the Hopfield model, but rather by a different set of time shifts (phases) between the periodic
activities of the neurons.
We study how the capacity of the network depends on the number of neurons, on the number of connections,
and on the topology of the network, that is
on the distribution of the connections between neighboring or distant neurons.

We consider a network composed by $N$ neurons, with $N(N-1)$ possible (directed) connections $J_{ij}$.
The activity of the neuron is represented by a time-dependent variable $\sigma_i(t)$, with $\sigma_i(t)=-1$ (silent neuron)
or $\sigma_i(t)=1$ (firing neuron).
Note that, in a coarse grained view, the variable $\sigma_i(t)$ may as well represent a group of neighboring neurons
with a highly correlated activity.

Not all the connections are present in the network. We put neurons randomly in a three-dimensional box
with periodic boundary conditions, with density equal to one.
For each neuron, we consider the sphere centered on it,  with a radius $R$ such that the sphere contains the $z$ nearest neurons.
Then we connect each neuron to $(1-\gamma)z$ neurons chosen randomly within the sphere of the neighbors, 
and $\gamma z$ neurons chosen randomly in the whole system.
In this way, we realize a network with a given mean connectivity $z$,
and a given fraction $\gamma$ of long range vs. short range connections.

\begin{figure}
\begin{center}
\includegraphics[width=.25\textwidth]{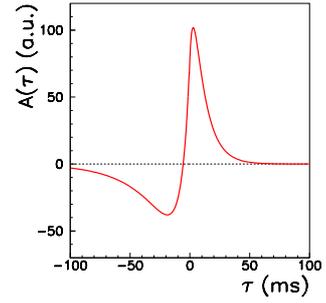}
\end{center}
\caption{%
The learning window $A(\tau)$ used in the learning rule to model STDP,
introduced and motivated by \cite{Abarbanel},
$A(\tau) = a_p e^{-\tau/T_p} - a_D e^{-\eta \tau/T_p}$ if $\tau>0$,
$A(\tau) = a_p e^{\eta\tau/T_D} - a_D e^{\tau/T_D}$ if $\tau<0$,
with $a_p = \gamma\,[1/T_p + \eta/T_D]^{-1}$,
$a_D = \gamma\,[\eta/T_p + 1/T_D]^{-1}$,
$T_p=10.2$ ms, $T_D=28.6$ ms, $\eta=4$, $\gamma=42$.
}
\label{fig_kernel}
\end{figure}

During the learning phase, we force the network to reproduce a number $P$ of spatio-temporal
periodic patterns, given by a specified function 
\begin{equation}
\sigma_i(t)=f(\omega^\mu t-\phi^\mu_i),
\end{equation}
where $\mu=1\ldots P$ is the index of the pattern, $\omega^\mu$ the angular velocity,
$\phi^\mu_i$ the phases of the neurons, that encode
the relative times at which neurons start to fire in the pattern $\mu$,
and the function $f(\phi)$ is periodic of period $2\pi$.
We consider the function
\begin{equation}
f(\phi)=\left\{\begin{array}{ll}
1&\mbox{if}\quad0<\phi\bmod 2\pi<\pi,\\
-1&\mbox{if}\quad\pi<\phi\bmod 2\pi<2\pi,
\end{array}\right.
\label{fperiod}
\end{equation}
with equal times of silent and firing state.
The connection $J_{ij}$ represents the strength of the synapse going from neuron $j$ to neuron $i$,
or in the coarse grained view the sum of strengths of synapses going from the group of neurons $j$ to the group of neurons $i$.
While in the Hopfield model the pattern to be stored $\sigma_i^\mu$
is static, and the learning rule is the outer product
$J_{ij} \propto \sigma_i^\mu \sigma_j^\mu$,
here the pattern to be stored $\sigma_i^\mu(t)$ is time dependent, and we formulate the change in the connections $J_{ij}$
in analogy with the STDP, as 
\begin{equation}
\delta J_{ij}=\int\limits_{0}^{t_{\text{max}}}\!\!dt\!\int\limits_{0}^{t_{\text{max}}}\!\!dt^\prime\,\, \sigma_i(t)
A(t-t^\prime) \sigma_j(t^\prime),
\end{equation}
where $A(\tau)$ is the learning function, and $[0, t_{\text{max}}]$
is the learning time interval in which the network is forced to reproduce the pattern $\mu$.
We use for the function $A(\tau)$ the one
introduced and motivated by \cite{Abarbanel},
with the parameters that fit the experimental data of \cite{biandpoo}
(see Fig.\ \ref{fig_kernel}).
In terms of Fourier components, this is written as
\begin{equation}
\delta J^\mu_{ij}=t_{\text{max}}\!\!\sum_{n=-\infty}^{\infty}
|\tilde{f}_n|^2\tilde{A}(n\omega^\mu)e^{in(\phi_i^\mu-\phi_j^\mu)},
\label{eqlearning}
\end{equation}
with $\tilde{f}_n=\frac{1}{2\pi}\int_0^{2\pi}\!d\phi\, f(\phi)e^{in\phi}$, and
$\tilde{A}(\omega)=\int\! dt\, A(t) e^{i\omega t}$. Due to the temporal asymmetry of $A(\tau)$,
the Fourier component $\tilde{A}(n\omega^\mu)$
has an imaginary part, and therefore $\delta J^\mu_{ij}\neq\delta J^\mu_{ji}$.
When we store multiple patterns $\mu=1,2,\ldots,P$,
the learned weights are the sum of the contributions from individual patterns. For the sake of simplicity, we consider
the same learning time $t_{\text{max}}$ and input frequency $\omega^\mu$ for all the encoded patterns. Moreover, for each
pattern $\mu$, we extract the phases $\phi_i^\mu$ randomly and uniformly from the interval $[0,2\pi)$.

\begin{figure}
\begin{center}
\setlength{\unitlength}{\textwidth}
\begin{picture}(0.45,0.25)
\put(-0.02,0){a)}
\put(-0.02,0){\includegraphics[width=.25\textwidth]{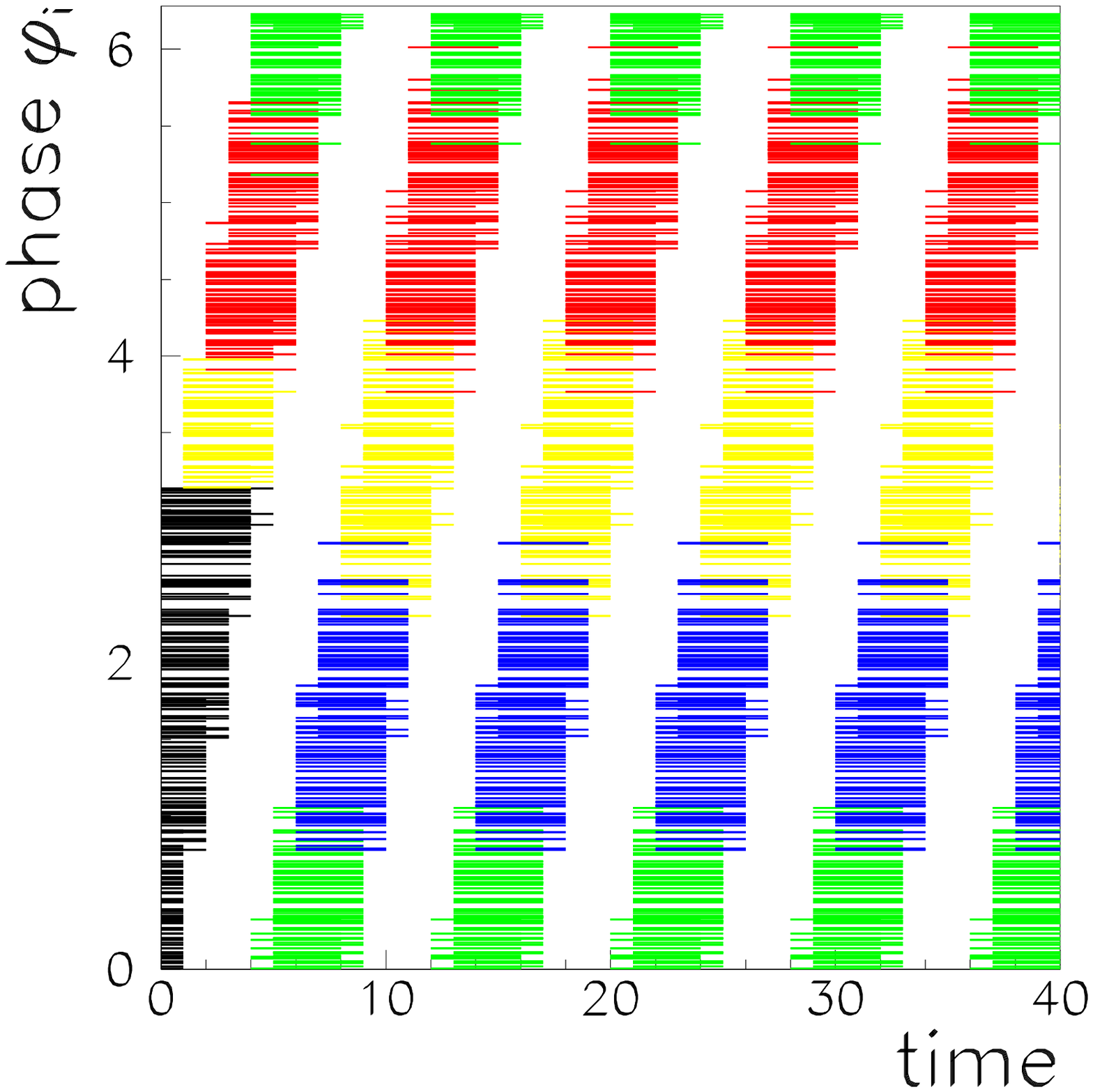}}
\put(0.22,0){b)}
\put(0.22,0){\includegraphics[width=.25\textwidth]{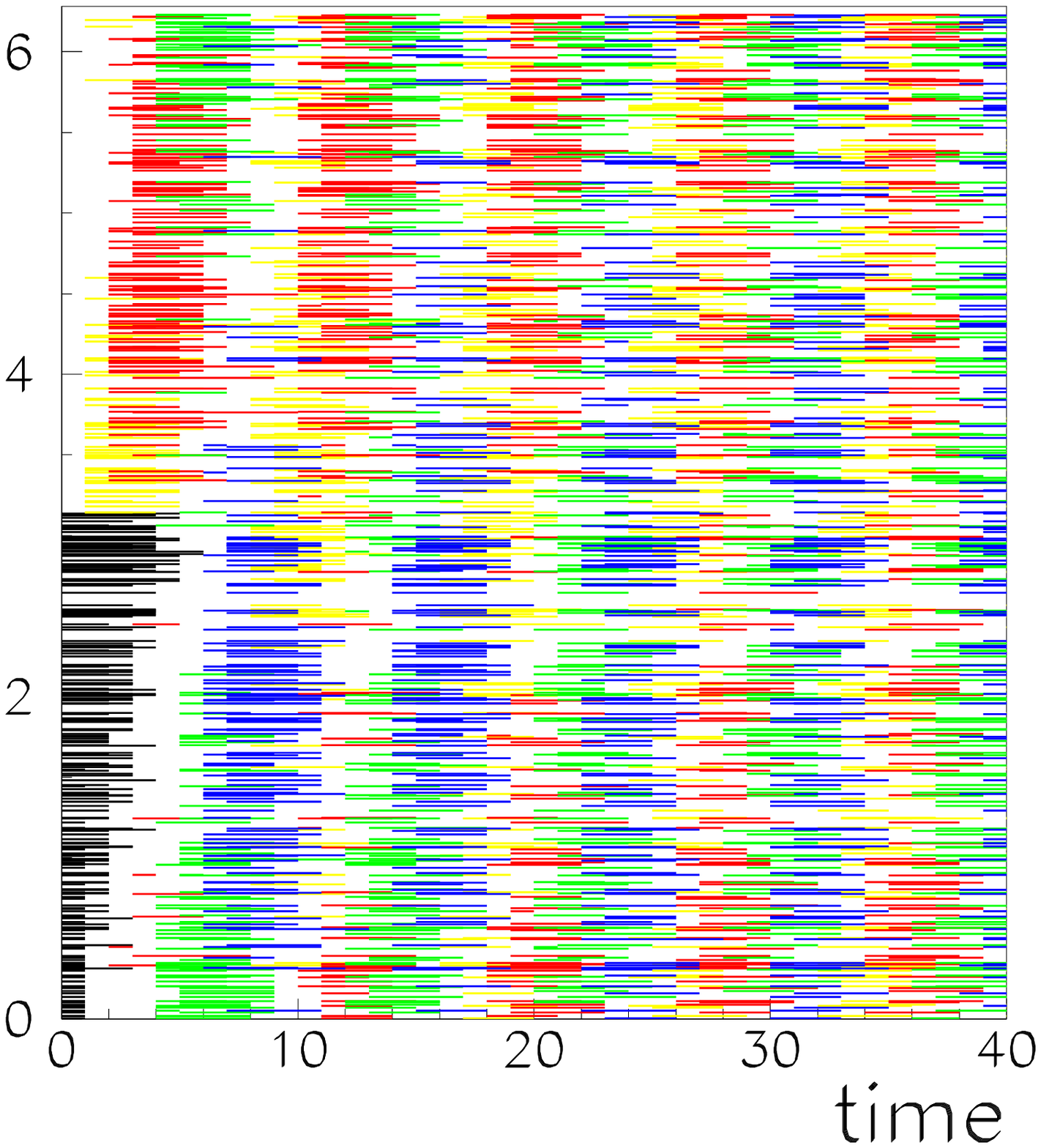}}
\end{picture}
\end{center}
\caption{%
a) Plot of the self-sustained dynamics of the fully connected network with $N=500$ and $P=5$.
The dynamics of the network, after a transient, is periodic of period $T=8\tau$.
Different colors represent the value of $\lfloor 4t_i/T\rfloor\bmod 4$,
where $t_i$ is time at which the neuron $i$ start to fire (black if neuron is firing at $t=0$).
In this case the times $t_i$ are highly correlated with the phases $\phi_i^1$, which means that the pattern
$\mu=1$ is well retrieved, and the overlap $m_1(t)$ is large.
b) The same network with $P=50$. In this case the pattern is not retrieved, and the overlap is of order $\sqrt{1/N}$.
}
\label{fig_dyn}
\end{figure}

After the learning phase, we initialize the network with a given initial condition $\sigma_i(0)$,
and perform a zero temperature spontaneous dynamics,
\begin{equation}
\sigma_i(t+\tau)=
\operatorname{sign}\left[\sum_{j\neq i}J_{ij}\sigma_j(t)\right]
\label{eqdynamics}
\end{equation}
with $\tau$ the unit step of time.
Due to the shape of the response function $\operatorname{sign}(x)$, if all the connections $J_{ij}$ are
multiplied by the same positive constant, the dynamics is unchanged. Therefore,
the learning time $t_{\text{max}}$ and the amplitude of the learning function $A(\tau)$ are immaterial.

In order to measure the similarity between the network activity during retrieval mode, and
the stored phase-coded pattern $\mu$, we define the overlap 
\begin{equation}
m^\mu(t) = \frac{1}{N}\sum_{i=1}^N \sigma_i(t) e^{i \phi_i^\mu}.
\end{equation}
When the retrieval is perfect, $\sigma_i(t)=f(\tilde\omega t-\phi^\mu_i)$,
with an output angular velocity $\tilde\omega$ that is in general different from the input $\omega^\mu$.
In this case $m^\mu(t)=\tilde{f}_1e^{i\tilde\omega t}$,
where $|\tilde{f}_1|\simeq 0.64$ for the function (\ref{fperiod}).
When the retrieval is not perfect, the modulus $|m^\mu(t)|$ after a transient goes to a constant value lower than
the maximal one.
Finally, if the network is not able to reproduce the pattern, the overlap becomes after a transient of order $1/\sqrt{N}$.
The modulus of $m^\mu(t)$ is therefore an order parameter which measure how much the network dynamics match the stored phase-coded pattern $\mu$. 
Note that the output frequency $\tilde\omega/2\pi$ depends on the input frequency $\omega^\mu/2\pi$,
and on the degree of asymmetry of the learning function $A(\tau)$. 
If $\tilde\omega$ is different from $\omega^\mu$, and $|m^\mu(t)|$ is high,
it means that the phase-coded stored pattern $\mu$ is replayed at
a frequency different from the one used to store it, i.e. at a different time scale, but with the same phase relationship. 
In Fig.\ \ref{fig_dyn}a we show the dynamics of a network of $N=500$ neurons, with all the $N(N-1)$ connections activated,
and with $P=5$ patterns encoded at input frequency $\omega^\mu/2\pi=10$ Hz.
The network is initialized with a high overlap with the pattern $\mu=1$, setting
\begin{equation}
\sigma_i(0)=\left\{%
\begin{array}{ll}
1&\mbox{if $0<\phi_i^1<\pi$},\\
-1\>&\mbox{if $\pi<\phi_i^1<2\pi$}.
\end{array}
\right.
\label{eqinit}
\end{equation}
Each segment represent a time interval in which the neuron $i$ is firing, that is $\sigma_i(t)=1$, with the neurons ordered on the vertical axis
by the value of the phase of the first pattern $\phi_i^1$.
In this case the pattern $\mu=1$ has been retrieved, and the overlap $m^1(t)$ has modulus $|m^1(t)|\simeq 0.63$ at long times.
In Fig.\ \ref{fig_dyn}b
we show the same network with $P=50$ patterns encoded. In this case the pattern is not retrieved, and the overlap has modulus $|m^1(t)|\simeq 0.07$.

\begin{figure}[t!]
\begin{center}
\setlength{\unitlength}{\textwidth}
\begin{picture}(0.45,0.25)
\put(-0.02,0){a)}
\put(-0.02,0){\includegraphics[width=.25\textwidth]{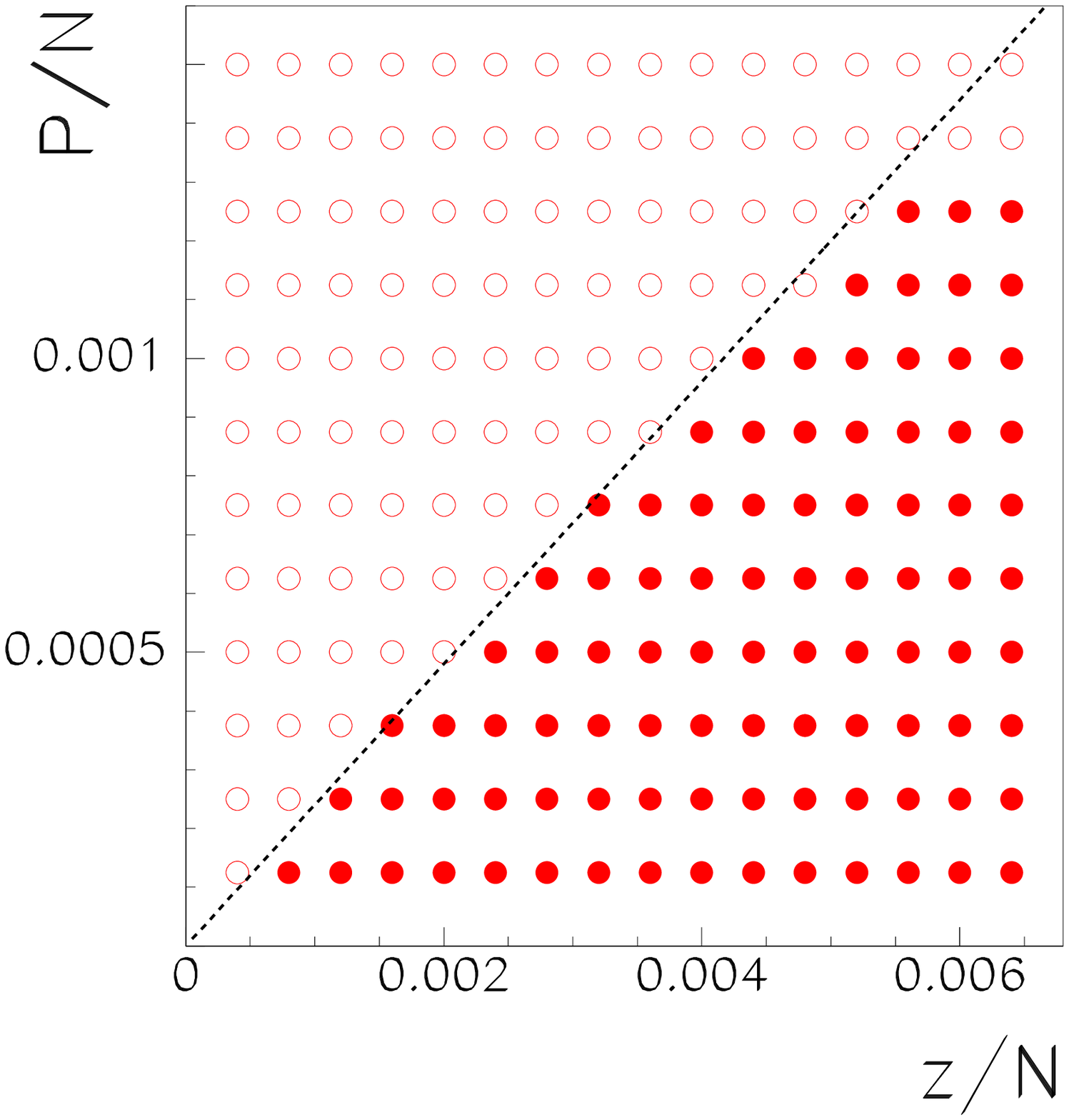}}
\put(0.22,0){b)}
\put(0.22,0){\includegraphics[width=.25\textwidth]{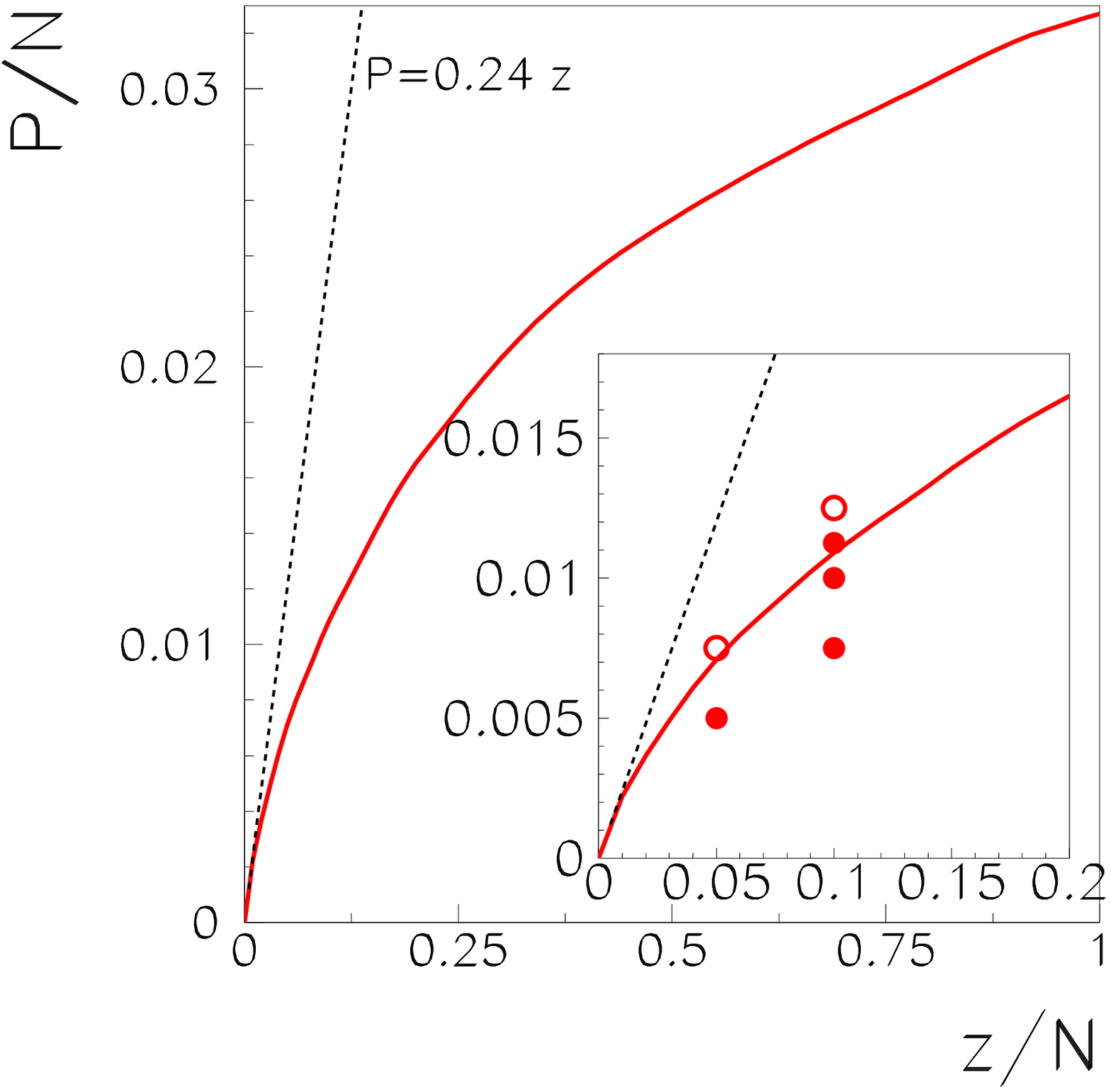}}
\end{picture}
\end{center}
\caption{%
a) Storage capacity of the network, as a function of the fraction of connections $z/N$ and the number of patterns per neuron $P/N$,
in the region of low connectivity, with $N=40000$ neurons.
Solid (empty) circles represent points where the network is (is not) able to retrieve patterns. 
The network with $N=4000$ gives the same result (within errors).
In this region of low $z/N$ the capacity is proportional to the number of connections.
b) Storage capacity in the entire range of $z/N$, with $N=4000$ neurons.
For clarity only a spline separating the encodable and the not-encodable region is shown.
Inset: comparison between the $N=4000$ and the $N=40000$ cases for two values of $z/N$.
The dashed line in all figures corresponds to $P=0.24\,z$.
}
\label{fig_sparse}
\end{figure}

We define the capacity $P_{\text{max}}$ of the  network as the maximum number of patterns encodable in the network, and retrievable
with an overlap greater than $0.2$.
We have first studied the capacity of the network as a function of the number of connections,
in the case of a fraction $\gamma=1$ of long range connections. As the probability to create a long range connections is
independent on the distance of the neurons, in this case the network is completely random.
We therefore consider a network of $N$ neurons, and create a connection from each neuron to a number $z$ of randomly chosen neurons,
with $0<z<N$.
We then look if the network is able to encode $P$ (random) pattern, and retrieve one of them subsequently.
We initialize the network with an high overlap with one of the patterns, as in Eq.\ (\ref{eqinit}), and simulate the dynamics 
in Eq.\ (\ref{eqdynamics}) up to a time $t=2\times 10^5$ in Monte Carlo steps, looking if the overlap with the chosen pattern
remains higher than $0.2$.  The experiment is repeated with three different sets of $P$ patterns.
In Fig.\ \ref{fig_sparse}a the result is shown for $N=40000$. Solid (empty) circles represent points where the network
was (was not) able to retrieve a pattern at least two times out of three. In practice, for all points considered, the network either was always
able to retrieve the pattern (three times out of three), or never (zero times out of three).
Furthermore, retrieved patterns had always an overlap greater than $0.36$, while not retrieved ones had an overlap lower than $0.08$.
Results for $N=4000$ were practically the same, with a small number of points near the separation between encodable and not encodable region
showing some fluctuations (retrieving the pattern one or two times out of three).
Note that, in the range of values of $z/N$ considered in Fig.\ \ref{fig_sparse}a, the maximum capacity $P_{\text{max}}$ is well described
by a linear function of the connectivity, $P_{\text{max}}=0.24\, z$.

In Fig.\ \ref{fig_sparse}b, we show the results for the entire range of $z/N$, from zero to one. When $z/N$ is of the order of unity,
the number of connections is of order $N^2$, so the simulation is too expensive for $N=40000$.
We therefore study sistematically only the network with $N=4000$, but compare it with the case $N=40000$ for two values of $z/N$,
finding a good agreement. For clarity we do not show the points simulated, but only a spline separating the encodable and the not-encodable region.
The dashed line shows the limit of the capacity for low connectivity $z/N$, $P_{\text{max}}=0.24\, z$.
It can be seen that, when the connectivity grows, there is a saturation effect, so that for $z\simeq N$ one finds
$P_{\text{max}}/N\simeq 0.032$, and therefore $P_{\text{max}}\simeq 0.032\,z$.

%
%
%
%
%
%
%
%

\begin{figure}
\begin{center}
\setlength{\unitlength}{\textwidth}
\begin{picture}(0.45,0.25)
\put(-0.02,0){a)}
\put(-0.02,0){\includegraphics[width=.25\textwidth]{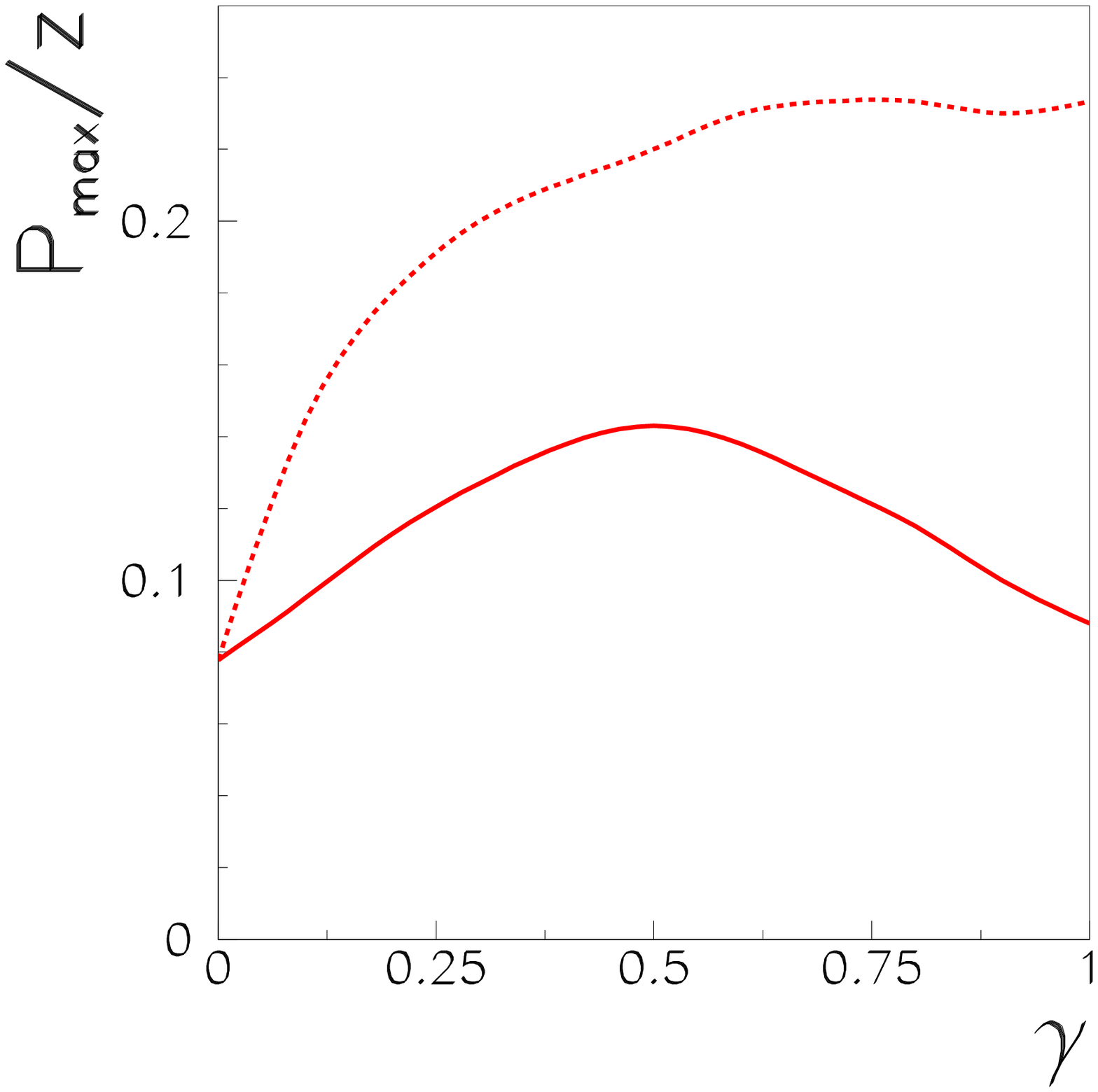}}
\put(0.22,0){b)}
\put(0.22,0){\includegraphics[width=.25\textwidth]{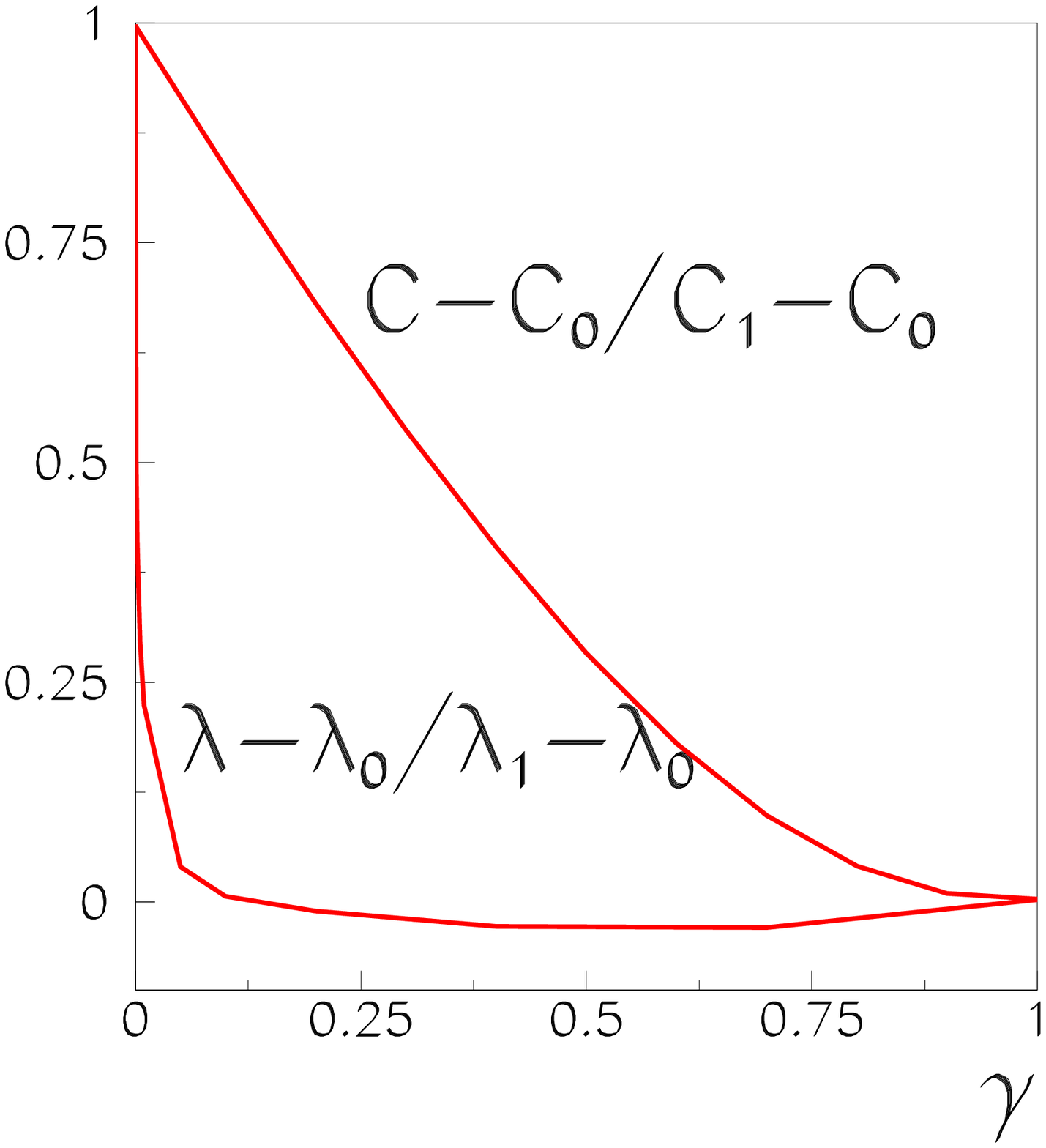}}
\end{picture}
\end{center}
\caption{%
a) Storage capacity as a function of the fraction of long range connections $\gamma$,
for a fixed number of connections per node (dashed line)
or for a fixed total cost of the connections (solid line).
b) Relative clustering coefficient and path length  as a function of $\gamma$.
}
\label{fig_3}
\end{figure}

To study the effects of short and long range connectivity, we have then analyzed the capacity of the network
in the case of a fraction $\gamma$ of long range connections, and $(1-\gamma)$ of short range connections.
The result is shown in Fig.\ \ref{fig_3}a (dashed line), for $N=40000$ and $z=180$ ($z/N=0.0045$). Note that in this case
the sphere containing the short range neurons has a radius $R\simeq 3.5$, while the side of the cubic box is $L\simeq 34$.
The storage capacity, that is the maximum encodable number of pattern such that the retrieval gives an overlap
greater than 0.2, goes from $P_{\text{max}}\simeq 0.11\,z$ for $\gamma=0$ (only local connections)
to $P_{\text{max}}\simeq 0.24\,z$ for $\gamma=1$ (random network).
This shows that the number of connections is not the only parameter that determines the capacity of the network.
Long range (random) connections
are more effective than short range ones in encoding patterns.

The topology of the biological networks has been shaped by the evolution,
as a compromise between the effectiveness in realizing complex tasks, and cost minimizations.
While long range connections are more effective than short range ones, they are of course more costly. The trade-off between
these two requirements will produce a network with a finite fraction of long range connections.
We therefore introduce a parameter $f$, that represents the cost of a long range connections  with respect to a short range one,
and we consider a network with
$(1-\gamma)z$ local connections, with neurons chosen randomly within those at distance lower than $R$,
and $\gamma z/f$ long range connections, with neurons chosen randomly in the whole system. Varying $\gamma$, in this case it is not
the connectivity that is constant, but the total cost of the connections.
The storage capacity as a function of $\gamma$ is shown in  Fig.\ \ref{fig_3}a (solid line), for the case $f=3$.
The maximum capacity of the network is realized with $\gamma\simeq 0.5$, that corresponds to about $25\%$ of long range connections
over the total of the connections.
To characterize the network, we have also calculated for the different values of $\gamma$ 
the clustering coefficient $C$,
defined as the probability that two neurons, that are connected
to a third neuron, are themselves connected,
and the mean path length $\lambda$, defined as
the  minimum number of connections needed to go from a node to another node, averaged over all pairs of nodes.
In Fig.\ref{fig_3}b, we show the relative clustering coefficient $(C-C_1)/(C_0 -C_1)$,
and path length  $(\lambda -\lambda_1)/(\lambda_0 -\lambda_1)$,
where the quantities with subscripts 0 and 1 refer to the cases $\gamma=0$ and $\gamma=1$ respectively.
Note that the clustering coefficient decreases much more slowly than the
path length, giving rise to a large region where the clustering coefficient is
not much lower than that for local connections only, while the path length is
almost as low as in the random network.
The network therefore, in the region of intermediate $\gamma$ that corresponds to optimal capacity,
is of a small-world type \cite{watts-strogats}.

In this paper we studied the storage and
recall of patterns in which information is encoded in the
phase-based timing of firing relative to the cycle. 
We proposed a STDP-based learning rule, and we analyzed the
its ability  
to memorize multiple phase-coded patterns, such that the
spontaneous dynamics of the network
selectively gives sustained activity which matches one of the stored
phase-coded patterns, depending on the initialization of the network.
Our work generalizes the Hopfield model, to dynamical periodic states,
characterized by the relative phases of the neurons.

We have studied the storage capacity for different degrees of
sparseness and topologies of the connections.
Changing the proportion $\gamma$ between short-range and long-range connections, we go from a three-dimensional
network with only nearest-neighbors connections ($\gamma=0$) to a random network ($\gamma=1$).
Small but finite values of $\gamma$ give a ``small world'' topology,
similar to that found in many areas
of nervous system.

We find that in the case of only short range connections  the capacity is lowest,
while in the case of only long range connections, that corresponds to a completely
random network, the capacity is highest.
Moreover, a small but finite fraction of long range connection is enough to enhance the capacity
highly, with respect to the short range case.
Therefore if the cost of the connections is taken in account, with long range connections more costly than short range
ones, than the optimal capacity will be given by a small fraction of long
range connections, that corresponds to a small-world topology.

This is in agreement with the observation that small-world attributes, with high clustering coefficient and short path length,
was found across multiple spatial scales of cortical organization \cite{sporns-chialvo,sporns1}.
This property, first found in C.\ elegans,
is highly conserved over different type of measurement, and
across different species, including cat, monkey and humans, for both functional and anatomical networks \cite{bullmore}.


\begin{thebibliography}{99}
\expandafter\ifx\csname url\endcsname\relax\def\url#1{\texttt{#1}}\fi

\bibitem{science}
\Name{Buzsaki G. \and Draguhn A.} \REVIEW{Science }{304}{2004}{1926}.

\bibitem{Singer99}
\Name{Singer W.} \REVIEW{Neuron }{24}{1999}{49}.

\bibitem{Fries2007}
\Name{Fries N., Nikoli\'c D. \and Singer W.} \REVIEW{Trends in Neurosciences
  }{30}{2007}{309}.

\bibitem{buz}
\Name{Pastalkova E., Itskov V., Amarasingham A. \and Buzsáki G.}
  \REVIEW{Science }{321}{2008}{1322}.

\bibitem{MillerPNAS}
\Name{Siegel M., Warden M.~R. \and Miller E.~K.} \REVIEW{Proc Natl Acad Sci USA
  }{106}{2009}{21341}.

\bibitem{8}
\Name{O'Keefe J. \and Recce M.~L.} \REVIEW{Hippocampus }{3}{1993}{317}.

\bibitem{burges}
\Name{Huxter J., Burgess N. \and O'Keefe J.} \REVIEW{Nature }{425}{2003}{828}.

\bibitem{zugaro-cn}
\Name{Geisler C., Robbe D., Zugaro M., Sirota A. \and Buzsaki G.} \REVIEW{Proc
  Natl Acad Sci USA }{104}{2007}{8149}.

\bibitem{odor}
\Name{Perez-Orive J., Mazor O., Turner G.~C., Cassenaer S., Wilson R.~I. \and
  Laurent G.} \REVIEW{Science }{297}{2002}{359}.

\bibitem{Montemurro}
\Name{Montemurro M.~A., Rasch M.~J., Murayama Y., Logothetis N.~K. \and Panzeri
  S.} \REVIEW{Current Biology }{18}{2008}{375}.

\bibitem{12}
\Name{Kayser C., Montemurro M.~A., Logothetis N.~K. \and Panzeri S.}
  \REVIEW{Neuron }{61}{2009}{597}.

\bibitem{panzieri}
\Name{Panzeri S., Brunel N., Logothetis N. \and Kayser C.} \REVIEW{Trends in
  Neurosciences }{33}{2010}{111}.

\bibitem{NC}
\Name{Scarpetta S., Zhaoping L. \and Hertz J.} \REVIEW{Neural Computation
  }{14}{2002}{2371}.

\bibitem{mate}
\Name{Lengyel M., Kwag J., Paulsen O. \and Dayan P.} \REVIEW{Nature
  Neuroscience }{8}{2005}{1677}.

\bibitem{Timme2006}
\Name{Memmesheimer R.~M. \and Timme M.} \REVIEW{Phys. Rev. Lett.
  }{97}{2006}{188101}.

\bibitem{PREYoshioka}
\Name{Yoshioka M., Scarpetta S. \and Marinaro M.} \REVIEW{Phys. Rev. E
  }{75}{2007}{051917}.

\bibitem{y}
\Name{Yoshioka M.} \REVIEW{Phys. Rev. Lett. }{102}{2009}{158102}.

\bibitem{Bush}
\Name{Bush D., Philippides A., Husbands P. \and O'Shea M.} \REVIEW{PLoS Comput
  Biol. }{6}{2010}{e1000839}.

\bibitem{masquelier}
\Name{Masquelier T., Hugues E., Deco G. \and Thorpe S.} \REVIEW{J. Neurosci.
  }{29}{2009}{13484}.

\bibitem{nature}
\Name{Flight M.~H.} \REVIEW{Nature Reviews Neuroscience }{10}{2010}{834}.

\bibitem{markram}
\Name{Markram H., Lubke J., Frotsher M. \and Sakmann B.} \REVIEW{Science
  }{275}{1997}{213}.

\bibitem{biandpoo}
\Name{Bi G.~Q. \and Poo M.~M.} \REVIEW{J. Neurosci. }{18}{1998}{10464}.

\bibitem{caporale}
\Name{Caporale N. \and Dan Y.} \REVIEW{Annu. Rev. Neurosci. }{31}{2008}{25}.

\bibitem{sjostrom}
\Name{Siostrom P.~J.} \REVIEW{Front. Syn. Neurosc. }{2}{2010}{4}.

\bibitem{17}
\Name{Koch C. \and Laurent G.} \REVIEW{Science }{284}{1999}{96}.

\bibitem{15}
\Name{Laughlin S.~B. \and Sejnowski T.~J.} \REVIEW{Science }{301}{2003}{1870}.

\bibitem{watts-strogats}
\Name{Watts D.~J. \and Strogatz S.~H.} \REVIEW{Nature }{393}{1998}{440}.

\bibitem{bullmore}
\Name{Bullmore E. \and Sporns O.} \REVIEW{Nat. Rev. Neurosci. }{20}{2009}{186}.

\bibitem{latora}
\Name{Latora V. \and Marchiori M.} \REVIEW{Eur. Phys. J. B }{32}{2003}{249}.

\bibitem{singer}
\Name{Yu S., Huang D., Singer W. \and Nikoli\'c D.} \REVIEW{Cerebral Cortex
  }{18}{2008}{2891}.

\bibitem{sporns1}
\Name{Sporns O. \and Zwi J.~D.} \REVIEW{Neuroinformatics }{2}{2004}{145}.

\bibitem{Abarbanel}
\Name{Abarbanel H., Huerta R. \and Rabinovich M.~I.} \REVIEW{Proc Natl Acad Sci
  USA }{99}{2002}{10132}.

\bibitem{sporns-chialvo}
\Name{Sporns O., Chialvo D., Kaiser M. \and Hilgetag C.} \REVIEW{Trends in
  Cognitive Science }{8}{2004}{418}.

\end{thebibliography}
\end{document}